\documentstyle{elsart}
\mathchardef\lrond="324C
\def\la{\mathrel{\mathchoice {\vcenter{\offinterlineskip\halign{\hfil
$\displaystyle##$\hfil\cr<\cr\sim\cr}}}
{\vcenter{\offinterlineskip\halign{\hfil$\textstyle##$\hfil\cr
<\cr\sim\cr}}}
{\vcenter{\offinterlineskip\halign{\hfil$\scriptstyle##$\hfil\cr
<\cr\sim\cr}}}
{\vcenter{\offinterlineskip\halign{\hfil$\scriptscriptstyle##$\hfil\cr
<\cr\sim\cr}}}}}

\begin{document}

\begin{frontmatter}
\title{Genus statistics on CMB polarization maps \\
and cosmological parameter degeneracy}

\author [iss]{L. Popa\thanksref{rfemail1}},
\author [iss]{P. Stefanescu\thanksref{rfemail2}},  \author [unifi]{R.
Fabbri\thanksref{rfemail3}}
\address [iss]{Institute for Space Sciences, Bucharest-Magurele,
R-76900, Romania}
\address [unifi]{Dipartimento di Fisica, Universit\`a di Firenze,  Via
S.~Marta,~3,  \\
I-50139 Firenze, Italy}
\thanks[rfemail1]{E-mail: lpopa@venus.ifa.ro}
\thanks[rfemail2]{E-mail: pstep@venus.ifa.ro}
\thanks[rfemail3]{E-mail: fabbri@fi.infn.it}

\begin{abstract}
We apply genus statistics to simulated CMB polarization maps,
constructed from secondary-ionization cosmological models
in experimental situations comparable to those of
forthcoming space experiments. We find that both the cosmic
baryon density and the spectral index of density perturbations
are strongly anticorrelated to the reionization redshift.
Using the Fisher matrix approach we show that
the accuracies in determining  the spectral index
and the optical depth to the reionization  epoch
are better in the case of genus statistics than in standard power spectrum
statistics.
\end{abstract}

\begin{keyword}
Cosmology: cosmic microwave background, large scale structure,
polarized radiation - Methodic: topology

PACS: 98.70.V, 98.80,11.27

\end{keyword}
\end{frontmatter}

\section{Introduction}

The cosmic microwave background (CMB) offers one of the best probes of the
early universe, enabling us to test the consistency of structure formation
models. The detection of the CMB anisotropies at large scales by the COBE
satellite (Smoot et al., 1992) as well as a number of new detections at
intermediate and small scales [see e.g., White et al. (1994), Scott et al.
(1995) and Bond (1996)] provided precious information to
constrain several cosmological parameters. In recent years there has been
an
increasing interest for the CMB polarization, leading to substantial
progress regarding the structure and solutions of the transport equations
(Ma \& Bertschinger, 1995; Hu et al., 1998),
numerical algorithms (Seljak \& Zaldarriaga, 1996; Zaldarriaga et al.,
1998),
and statistical
descriptors (Melchiorri \& Vittorio, 1996;
Seljak, 1997; Ng \& Liu, 1998; Naselsky \& Novikov, 1998).

As CMB  polarization is produced when the anisotropic radiation
possessing  non-zero quadrupole moment is scattered by free electrons
via Thomson scattering (Rees, 1968), its
magnitude,
spatial distribution and topological properties can provide
information that is complementary to that obtained from the anisotropy
alone.
Unlike the temperature fluctuations which may evolve between the last
scattering and today,  the CMB polarization probes the epoch of the last
scattering directly. Also, different sources of
temperature anisotropies (scalar and tensorial) give different signatures
in the polarization power spectra.   Therefore polarization is a powerful
tool for
reconstructing the  sources of anisotropy.   A full description of
polarized
radiation requires
the introduction of  three additional power spectra,  defining the
``electric" and ``magnetic" components of the polarization vector pattern
and the
cross correlation with the temperature anisotropies
(Seljak \& Zaldarriaga, 1996; Efstathiou \& Bond 1998).
Including this additional information one can distinguish
among physical processes generating  the temperature power spectrum,  and
hence better constrain  cosmological models  (Zaldarriaga, Spergel \&
Seljak,
1997).

The current measurements of CMB temperature anisotropy already permit
to place significant constraints on certain cosmological parameters.
These parameters include the amplitudes and the spectral indices of
scalar and tensor perturbations (Knox \& Turner, 1994; Crittenden,
Coulson \& Turok, 1995; Knox, 1995;  Lidsey et al., 1997;
Souradeep et al., 1998; Copeland et al., 1998),
the various components of the mass density of the Universe and
the Hubble constant (Jungman et al., 1996b; Lineweaver et al., 1997, 1998;
Bond, Efstahiou \& Tegmark, 1997; Zaldarriaga, Spergel \& Seljak, 1997;
Bond, Jaffe \& Knox, 1998; Zaldarriaga, 1998; Hancock et al., 1998;
Webster et al., 1998; Bartlett et
al., 1998a, 1998b).
The role of the CMB polarization in the determination of the cosmological
parameters has been discussed in several works
(Zaldarriaga, 1997,1998;
Seljak, 1997; Kamionkowski et al. 1997a;
Efstathiou \& Bond, 1998).
Their results show that, although  polarization can
improve the accuracy of many cosmological parameters only by a modest
amount,
there are  two important exceptions,
regarding the discrimination of scalar and tensor modes and
the determination of  the reionization redshift of the intergalactic
medium.
Some degeneracies among cosmological parameters
which are expected to arise in future measurements of CMB anisotropies will
be broken by the joint exploitation of polarization and  temperature
anisotropy
(Kinney, 1998), and others by the simultaneous use of some more
astronomical data (Efstathiou \& Bond, 1998).

The main limitation for the cosmological use of polarization is that it is
predicted to be small. The best experimental upper limit now available,
being $P \la 16$ $\mu $K at a scale of 1.4$^{\circ }$
(Netterfield et al., 1995),
is not very stringent from a theoretical point of view. However, a number
of
experiments which are being planned both at ground [POLAR (Keating et al.,
1998), MITO (de Petris, 1998)] and from space [MAP (Bennet et al., 1996),
SPOrt (Cortiglioni et al., 1997), PLANCK (Bersanelli et al., 1996)]
are expected to improve the
situation quite substantially.
In view of the best exploitation of such forthcoming experiments (mainly,
of
MAP and PLANCK) detailed work has been performed on statistical descriptors
of polarization. However, previous full-sky statistical studies of the
polarization are generally based on the power spectrum estimators and
cross-correlations in Fourier (Seljak \& Zaldarriaga, 1997;
Kamionkowski et al., 1997a; Zaldarriaga \& Seljak, 1997)
and real space (Ng \& Liu, 1998). A larger variety of methods have been
proposed for studies
of temperature maps:
These include the analysis of 2-point correlation functions (Hinshaw et
al., 1996),
power spectra in terms of spherical harmonics  modified for Galactic cut
(Wright et al.,
1994) and in terms of  Karhunen-Lo\'eve   eigenmodes (Bunn \& Sugiyama,
1995), eventually
including data compression  (Tegmark et al., 1997), and topological methods
(Torres
et al., 1995). As to the latter methods, in the case of polarization
fields, only  general
topological properties  have been treated   in
the recent works of Naselsky \& Novikov
(1998) and Dolgov et al. (1998).

In this paper we analyze the topological structure of the CMB polarized
field employing genus statistics in order to study the degeneracies among
parameters of cosmological models. Our choice is motivated by the fact that
genus is a locally invariant statistical estimator
(Bond \& Efstathiou, 1987; Gott et al., 1990; Torres et al., 1995;
Schmalzing \& Buchert, 1997), in the sense that an incomplete and
non-uniform sky coverage leaves this quantity unchanged. In order to
consider a realistic experimental situation, we perform simulations
adopting
the large but incomplete sky coverage consistent with the environment on
board the International Space Station Alpha (ISSA) and actually planned for
SPOrt. We also adopt the angular resolution $FWHM=7^{\circ }$ competing to
POLAR and SPOrt, and to previous topological analyses of COBE's
anisotropy field (Fabbri \& Torres, 1996).

The above angular scale is very interesting for cosmological models with a
secondary ionization (Seljak \& Zaldarriaga, 1997; Kamionkowski et al.,
1997b; Zaldarriaga \& Seljak, 1997), which
predict the existence of a broad peak in the polarization power spectrum at
low order multipoles $(l
\la 30)$ that is not present in the anisotropy power spectrum
(Zaldarriaga, 1997). In flat-space models, the peak position scales as $%
l\propto \tau _{ri}^{1/3}(h\Omega _bx_e)^{-1/3}$, where $\tau _{ri}$ is the
reionization optical depth, $\Omega _b$ the density parameter of the
baryonic matter, $h$ the reduced Hubble constant and $x_e$ the ionization
fraction. In general, large-scale polarization is enhanced in reionization
models (Bond \& Efstathiou, 1987; Zaldarriaga \& Harari, 1995;
Ng \& Ng, 1995), the main peak amplitude being
roughly proportional to $\tau _{ri}$ for $\tau _{ri}
\la 1$ (and also dependent on the spectral index of the primordial
power spectrum) (Zaldarriaga, 1997). For many of our model simulations we
assume a rather strong reheating, $\tau _{ri}\simeq 1,$ which is
marginally consistent
with experimental limits
(de Bernardis et al., 1997). In such a case we show that quite
significant results can be obtained from genus analysis in experiments with
a pixel sensitivity of 1 $\mu $K, which is quite consistent with
MAP's expected noise considering a $7^{\circ }$ beam averaging. For
weaker reionizations, which are more likely on experimental (de Bernardis
et
al., 1997; Fabbri, 1998)
and theoretical (Haiman \& Loeb, 1997a, 1997b)
grounds, similar results can be obtained
just scaling the noise level; as a matter of facts, the genus does not
depend on the absolute amplitudes of signal and noise, but only on their
ratio. Thus for instance, similar results for models with $\tau _{ri}\simeq
0.1$ can be obtained at the sensitivity level expected for PLANCK.

The polarization genus test of cosmological models is affected by its own
parameter degeneracies. Here we consider standard CDM models, assuming
scalar modes (with primordial spectral slope $n_s)$ and adiabatic initial
conditions. Spanning the four-dimensional parameter space $%
s_4=(z_{ri},n_s,\Omega _b,x_e),$ we find that $n_s$, $\Omega _b$ and $x_e$
are strongly anticorrelated to $z_{ri}$. The high-likelihood regions in
parameter space are found to be quite narrow from our simulations. Fixing
$%
z_{ri}=100$ for both input and target models, the other parameters could be
determined to a high accuracy for a 1 $\mu $K detector noise; for instance,
$\delta
n_s\simeq \pm 0.05$ and $\delta \Omega _b\simeq \pm 0.008$ at 68\% CL. (All
errors
here and henceforth obviously include cosmic variance.) We
conclude that a topological analysis of polarization can contribute to
obtain a very accurate determination of parameters when combined with some
other
test.

The most  common way   to evaluate  cosmological parameters
 is to use the power spectrum estimates in connection with
the standard likelihood function
(see e.g. Bartlett et al.,1998b; Efstathiou \& Bond, 1998; Zaldarriaga
1998).
It is thereby worth making a detailed comparison of the advantages and
disavantages of the genus technique versus the power spectrum technique.
To this purpose, we computed formal errors by means of  the Fisher
information
matrix method (e.g., Bond et al., 1997; Tegmark et al., 1998).
This method allows us to quickly compare the relative accuracies of
different
techniques, although it cannot be used for a very accurate determination of
likelihood
contours in parameter space. We performed calculations for the experimental

configurations pertaining to MAP, SPOrt and PLANCK, and
found that the genus analysis is more efficient than the power spectrum
analysis
for determining  $\tau_{ri}$ and  $n_s$, while the opposite is generally
true for $ \Omega _b$
(with the exception of the SPOrt configuration).
We also found that combining anisotropy and polarization data
(for instance, performing the genus analysis on both maps) typically
reduces the errors
by a factor $\sim 3$.

The plan of the paper is the following.
In the next  Section we present the formalism and method used for
Monte-Carlo simulations of the Stokes parameter $Q$ and $U$. The analysis
of the
polarization maps in terms of genus statistics is given in Section 3,  and
Section 4 presents the main results concerning the confidence regions of
cosmological parameters obtained from simulations.
In Section 5 we compare the accuracies in the estimates  of
the cosmological parameters  within
the Fisher information matrix
approach.

\section{Polarization maps}

The all-sky polarization maps were obtained following the formalism
described in
 Zaldarriaga \& Seljak (1997), by expanding the Stokes parameters $Q$ and
$U$  in spin-weighted spherical harmonics with random amplitudes
[cfr. also
Sazhin  \& Benitez (1995)].

\subsection{Stokes parameters}

The CMB polarization field can be described by a $2\times 2$ temperature
perturbation tensor $T_{ij}$, in terms of which the Stokes parameters $Q$
and $U$ and the temperature anisotropy $T$ are given by
(Kosowsky, 1996; Bondi \& Efstathiou, 1987) $Q=(T_{11}-T_{22})/4,$
$U=(T_{12})/2,$ $%
T=(T_{11}+T_{22})/4.$ (Circular polarization is not necessary because it
cannot be generated by Thomson scattering.) The combinations $Q\pm iU$ are
quantities of spin $\pm 2$ and for a given direction in the sky ${\hat n}$
can be expanded in spin-weighted spherical harmonics $_{\pm 2}Y_{lm}$,
\begin{equation}
(Q\pm iU)(\hat n)=\sum_{l,m}a_{\pm 2,lm}\;_{\pm 2}Y_{lm}(\hat n),
\label{spin2}
\end{equation}
while the temperature anisotropy is
\begin{equation}
T(\hat n)=\sum_{l,m}a_{T,lm}Y_{lm}(\hat n).  \label{spin0}
\end{equation}
The linear combinations
\begin{equation}
a_{E,lm}=-(a_{2,lm}+a_{-2,lm})/2,
\end{equation}
\begin{equation}
a_{B,lm}=(a_{2,lm}-a_{-2,lm})/2i
\end{equation}
can be directly related with power spectra $C_{Xl}$, which in the Gaussian
theory are rotational invariant quantities:
\begin{equation}
<a_{X,l^{^{\prime }}m^{^{\prime }}}^{*},a_{X,lm}>=\delta _{ll^{^{\prime
}}}\delta _{mm^{^{\prime }}}C_{Xl},  \label{power}
\end{equation}
\begin{equation}
<a_{T,l^{^{\prime }}m^{^{\prime }}}^{*},a_{E,lm}>=\delta _{ll^{^{\prime
}}}\delta _{mm^{^{\prime }}}C_{Cl},
\end{equation}
Here $X$ stands for $T$,$E$ or $B$ denoting the temperature,
electric-parity
and magnetic-parity polarization modes respectively, and $C_{Cl}$ describes
the $E$-$T$ cross correlation. (The $B$-$T$ and $B$-$E$ cross correlations
vanish because of the opposite parities.) From Eq. (\ref{spin2}) we obtain
$%
Q(\hat n)$ and $U(\hat n)$ (Zaldarriaga, 1997)
\begin{equation}
Q(\hat n)=-\sum_{lm}[a_{E,lm}X_{1,lm}(\hat n)+ia_{B,lm}X_{2,lm}(\hat n)],
\label{q}
\end{equation}
\begin{equation}
U(\hat n)=-\sum_{lm}[a_{B,lm}X_{1,lm}(\hat n)-ia_{E,lm}X_{2,lm}(\hat n)],
\label{u}
\end{equation}
with
\begin{eqnarray}
X_{1,lm}(\hat n) &=&(_2Y_{lm}+_{-2}Y_{lm})/2=\sqrt{(2l+1)/4\pi }%
F_{1,lm}(\theta )e^{im\phi },  \label{real1} \\
X_{2,lm}(\hat n) &=&(_2Y_{lm}-_{-2}Y_{lm})/2=\sqrt{(2l+1)/4\pi }%
F_{2,lm}(\theta )e^{im\phi },  \label{real2}
\end{eqnarray}
where the functions $F_{1,lm}$ and $F_{2,lm}$ can be calculated in terms of
Legendre polynomials (Zaldarriaga, 1997; Kamionkowski et al., 1997a).
The conditions
\begin{eqnarray}
X_{1,lm}^{*} &=&X_{1,l-m},\;X_{2,lm}^{*}=-X_{2,l-m},  \nonumber \\
a_{E,lm} &=&a_{E,l-m}^{*},\;a_{B,lm}=a_{B,l-m}^{*},  \label{cond2}
\end{eqnarray}
make $Q$ and $U$ real.

For the computation of $Q$ and $U$ we need sets of random realizations of
$%
a_{X,lm}$ consistent with assigned power spectra $C_{Xl}$ and
with the Gaussian distributions accounting for cosmic variance.

\subsection{Monte-Carlo simulations of the Stokes parameters}

A given cosmological model only provides the power spectra. We considered
standard CDM models without and with reionization (denoted by sCDM and
srCDM, respectively), assuming only scalar modes with primordial spectral
slope $n_s$ and adiabatic initial conditions. We spanned the
four-dimensional parameter space
\begin{equation}
s_4=(z_{ri},n_s,\Omega _b,x_e)   \label{s4}
\end{equation}
fixing $h=0.5$ and $Y_{p}=0.24$.
As input models we assumed either sCDM or
srCDM with $z_{ri}=100,$ but for the Monte-Carlo simulated grids of target
CDM models we investigated a fully 4-dimensional volume of $s_4$. The four
grid steps were 20, 0.05, 0.01 and 0.1, respectively, and interpolation was
further performed for computation of the confidence regions (see next
Section).

All the relevant power spectra $C_{T,l}$, $C_{E,l}$ and $C_{C,l}$ were
obtained using the CMBFAST code developed by Seljak and Zaldarriaga
(1996).

For each realization the coefficients $a_{E,lm}$ were obtained following
the
procedure given in Zaldarriaga \& Seljak (1997): For each multipole $l$ we
diagonalize the correlation matrix: $M_l=\left(
\begin{array}{c}
C_{Tl}\;C_{Cl} \\
C_{Cl}\;C_{El}
\end{array}
\right) ,$ then we multiply the square roots of the eigenvalues of $M_l$ by
a pair of Gaussian random numbers and rotate back to the original frame.
Following this procedure we obtained random realizations of $a_{E,lm}$
satisfying the correct correlation properties.

We constructed simulated maps of the Stokes parameters adopting a
pixelization scheme of the ``igloo'' type (Crittenden \& Turok, 1998).
The sky region seen by a 7$^{\circ }$ beam of an experiment on ISSA ($%
-51^{\circ }.6\leq \delta\leq 51^{\circ }.6$) was divided into rows with
edges
of constant latitude; each row was cut by constant-longitude lines, the
angular distance between two neighbor pixels of constant latitude being $%
\Delta \alpha =FWHM/\cos \delta$. Although the pixels obtained have unequal
trapezoidal shapes (becoming nearly rectangular only close to the Galactic
plane), we have the following
advantage, that the pixel edges defined by the
spherical-coordinate frame allow a fast integration of the spin-weighted
spherical harmonics. For each pixel we calculate $Q(\theta _i,\phi _i)$ and
$%
U(\theta _i,\phi _i)$ (with $\theta _i=\pi /2-\delta$ and $\phi
_i=\alpha$), taking
$l_{max}=30.$ Equations (\ref{q}) and (\ref{u}) with the conditions (\ref
{cond2}), including the finite beamwidth and adding the detector  noise
contribution, become
\begin{eqnarray}
Q(\theta _i,\phi _i)
&=&-\sum_{l=2}^{30}%
\sum_{m=0}^l[B_l(a_{E,lm}X_{1,lm}+a_{E,lm}^{*}X_{1,lm}^{*})]+N_i,
\nonumber
\\
U(\theta _i,\phi _i)
&=&-i\sum_{l=2}^{30}%
\sum_{m=0}^l[B_l(a_{E,lm}X_{2,lm}-a_{E,lm}^{*}X_{2,lm}^{*})]+N_i,
\label{uwnois}
\end{eqnarray}
where $B_l=\exp $ $\left[ -\sigma ^2l(l+1)/2\right] $ is the Gaussian
window
function of the beam, with $\sigma =0.425\times FWHM$ and $FWHM=7^{\circ
},$
and $N_i$ is a random realization of the noise per pixel. The {\em rms}
noise depends on the instrument sensitivity and the observing time for each
pixel. The average value of the {\em rms} detector noise per pixel used in
the simulations
was $1$ $\mu $K.

\section{Genus analysis}

The specific signature of the CMB polarization field $P=\sqrt{Q^2+U^2}$
obtained for different underlying cosmological models was analyzed using
the
Euler characteristic of the field, equivalent to genus per unit area, under
the assumption of random Gaussian primordial density perturbations. We
determine the integrated genus per unit area $G(p)$ above some threshold
$p$
(Naselsky \& Novikov, 1998; Dolgov et al., 1998) as
\begin{equation}
G(p)=N_{max}+N_{min}-N_{sad},  \label{genus}
\end{equation}
where $N_{max}$, $N_{min}$ and $N_{sad}$ are the number densities of
maxima,
minima and saddle points of $P$, respectively, above the threshold. The
expectation values in terms of the curvature we have:
\begin{equation}
<G(p)>=\frac 1{4\pi }\left( \frac{\sigma _1}{\sigma _0}\right)
^2(p^2-1)e^{-%
\frac{p^2}2},  \label{genusexpect}
\end{equation}
where $\sigma _0$ and $\sigma _1$ are the spectral parameters (Bond
\& Efstathiou, 1987b)
\begin{equation}
<Q^2>=<U^2>=\sigma _0^2,
\end{equation}
\begin{equation}
<Q_iQ_j>=<U_iU_j>=\delta _{ij}\frac{\sigma _1^2}2,
\end{equation}
and we set $Q_i=\partial Q/\partial x_i$ and $U_i=\partial U/\partial x_i$.
For each random realization of the Stokes parameters we calculated the
genus
distribution for 30 values of $p$. Figure 1 
\begin{figure}
\caption{The  integrated genus per unit area for some reionization
models. From top to bottom $z_{ri} = 200,150,100,50$, and $0$.
For all of the curves we take $\Omega_{b}=0.05$, $h=0.5$,
$Y_{p}=0.24$, $x_{e}=1$ and $n_{s}=1$.}
\end{figure}

\begin{figure}
\caption{ The  $\chi^{2}$ distributions obtained taking as a target model
srCDM, and as input models srCDM (continuous line) and sCDM
(dashed line).}
\end{figure}

presents the integrated genus distributions $<G(p)>$ obtained averaging
over
a set of 400 realizations of the Stokes parameters for different underlying
reionization models. For such models we take $\Omega _b=0.05$, $h=0.5$,
$Y_p=0.24$, $x_e=1$ and $n_s=1$. For each set of Monte-Carlo realizations
obtained for a given target model we calculate the $\chi ^2$ estimator
defined as
\begin{eqnarray}
\chi ^2 &=&\sum_{i=1}^{30}\sum_{j=1}^{30}(<G^{tg}(p_i)>-<G^{in}(p_i)>)
\lambda _{ij}^{-1}
\nonumber \\
\;\;\;\;\;\;\;&( &< G^{tg}(p_j)>-<G^{in}(p_j)>),
\label{chi2}
\end{eqnarray}
where $<G^{tg}(p)>$ and $<G^{in}(p)>$ are the ensemble-averaged integrated
genus for Monte-Carlo realizations of the target and input model
respectively, and $\lambda _{ij}$ is the covariance matrix of the
Monte-Carlo realizations of the target model:
\begin{eqnarray}
\lambda _{ij} &=&\frac
1{N_{realiz}}\sum_{k=1}^{N_{realiz}}(G^k(p_i)-<G(p_i)>)  \nonumber \\
\;\;\;\;\;&(&G^k(p_j)-  < G(p_j)>).  \label{lambdamatrix}
\end{eqnarray}
Figure 2 presents two $\chi ^2$ distributions obtained taking srCDM as
target model, and srCDM and sCDM as input models.

\section{The confidence regions}

The confidence regions in four-dimensional parameter space
$s_4=(z_{ri},n_s,\Omega_b,x_e)$ was obtained as constant $\chi ^2$
boundaries at $\chi _{min}^2+\Delta \chi _\nu ^2$ (Press et al. 1992),
for $\nu
=4 $ and $\Delta \chi _\nu ^2=4.72$ and 7.78 at $1$-$\sigma $ and
$2$-$\sigma $
level respectively.

\begin{figure}
\caption{The confidence contour at 1 $\sigma$ level (continuous lines)
in the $z_{ri}$-$n_{s}$ plane for target reionized CDM models
in four parameter space, when the input model was srCDM. The dashed
line gives the best-fit curve.}
\end{figure}
\begin{figure}
\caption{The confidence contour at 1 $\sigma$ level (continuous lines)
in the $z_{ri}$-$\Omega_{b}$ plane for target reionized CDM models
in four parameter space, when the input model was srCDM. The dashed
line gives the best-fit curve.}
\end{figure}

Figure 3 presents $1$-$\sigma $ confidence interval obtained for the
spectral index
of the scalar modes $n_s$ as a function of reionization redshift $z_{ri}$.
The contour is obtained taking as input model srCDM with $%
s_4=(100,1,0.05,1)$. We found that $n_s$ and $z_{ri}$ are highly
anticorrelated. The statistical fit for our range of parameters gives
\begin{equation}
n_s-1=-(7.12\pm 0.58)\times 10^{-5}z_{ri}^{3/2}  \label{fitnz}
\end{equation}
at $95$ $\%$ CL. The reason for this result is that the polarization
amplitude at the peak of the power spectrum is roughly proportional to the
reionization optical depth
\begin{equation}
\tau _{ri}\approx 3.8 \times 10^{-2}\left( x_e\Omega _bh\right) \Omega
^{-1/2}(1+z_{ri})^{3/2},  \label{tau}
\end{equation}
and also increases with the spectral index $n_s$ (i.e., enhancing small
scale perturbations). Thus the effect of decreasing $\tau _{ri}$ is
compensated by changing the spectral index $n_s$ by an amount proportional
to $\tau _{ri}\propto z_{ri}^{3/2}$.

It is worth noticing that genus does not depend on the absolute amplitude
of
the polarization field, but only on its angular power spectrum. Thus for
ideal, zero-noise experiments we do not expect Eq. (\ref{fitnz}) to be
valid
any more. However, since the detector noise has a quite different spectrum
from the cosmological signal, the signal amplitude is important in
practice.
Equation (\ref{fitnz}) should be regarded as an analog of equations
describing the $n_s-Q_{{\rm rms-PS}}$ anticorrelation found from COBE-DMR
anisotropy data, including the angular correlation function (Seljak
\& Bertschinger, 1993)
and topological analyses (Fabbri \& Torres, 1996).

Figure 4 presents the confidence contours in $\Omega _b$-$z_{ri}$ plane at
68 $\%$ CL, obtained in the four-dimensional parameter space for $h=0.5$.
Here again we find a clear anticorrelation described by
\begin{equation}
\Omega _b=(7.55\pm 0.36)\times 10^{-2}-(2.33\pm 0.12)\times
10^{-5}z_{ri}{}^{3/2}  \label{obz}
\end{equation}
with errors at $95 \%$ CL, and an analogous anticorrelation is found for
the couple $x_e$-$z_{ri}.$ Equation (\ref{obz}), too, is qualitatively
interpreted by means of Eq.  (\ref{tau}). It should be noted however that
$%
\tau _{ri}$ is not constant along the curve defined by (\ref{obz}). As a
matter of facts, although $\tau _{ri}$ is the most important parameter, the
polarization field does also depend on other parameters.

If $z_{ri}{}$ is fixed, then the other parameters are determined to a high
accuracy: We get $\delta n_s\simeq \pm 0.05$ and $\delta \Omega _b\simeq
\pm
0.008$ at 95\% CL. Setting $z_{ri}{}$ equal to the input value, Eqs. (\ref
{fitnz}) and (\ref{obz}) give displacements $\Delta n_s\simeq -0.07$ and $%
\Delta \Omega _b\simeq 0.002$ with respect to the input values. These
numbers can be interpreted as estimates of bias; however, since they are
comparable to errors and derived from the analytic best-fit curves, they
should probably be regarded as upper limits on bias.

The above accuracies refer to models with a rather strong reheating, $\tau
_{ri}\simeq 1,$ which provide polarized signals of several $\mu $K, greater
than the assumed pixel sensitivity of 1 $\mu $K. In an analysis of this
kind, such a sensitivity should be warranted over a large sky coverage
(about 80\% of the full solid angle in our simulations, i.e. over $\sim
600$
pixels). This requirement is somewhat beyond several of the currently
planned experiments. However, MAP's expected sensitivity of about 20
$\mu $K at a scale of 0.3$^{\circ }$ scales just to 1 $\mu $K considering a
$%
7^{\circ }$ beam averaging. For weaker reionizations, we should consider
that results rather similar to those presented here can be obtained just
scaling the noise level, since genus does not depend on the absolute
amplitudes of signal and noise, but only on their ratio. An optical depth
$%
\tau _{ri}=0.1$ is quite realistic in the light of ionizing source models
(Haiman \& Loeb, 1997a, 1997b) and an analysis of anisotropy data
(de Bernardis et al., 1997); in that case
the noise level that we request on a 7$^{\circ }$ scale is of order 0.1
$\mu
$K in order to get similar accuracies. This is consistent with
PLANCK's planned sensitivity at 100 GHz.
\section{Accuracy on the estimates of cosmological parameters}
In this section we investigate how measurements of the CMB
anisotropies alone and anisotropies plus polarization
can constrain the relevant cosmological parameters.
The errors on the cosmological parameter estimates
have few dominant components:
\begin{itemize}
\item A nearly exact or ``geometrical" degeneracy, that
leads to nearly identical power spectra
(Bond, Eftathiou \& Tegmark 1997; Efstathiou \& Bond 1998) provided we have
identical matter content,  primordial power spectra and
angular size distance to the the last scattering surface.
\item The cosmic variance, that results from comparing a theoretical
statistical distribution of observables with a finite distribution
represented by the data.
Cutting out parts of the sky,
such as the Galactic plane, increases the cosmic variance by a factor
approximately inversely proportional to the fraction of the sky sampled.
\item The method used ($\chi^{2}$, maximum likelihood,
the Fisher information matrix approximation) can also bias
estimates of the cosmological parameters.
\end{itemize}
The standard way to estimate parameters
is the maximization of the likelihood function.
If the likelihood function $\lrond$ for a given particular
data set  ${\bf D}$ is a multivariate Gaussian  in ${\bf D}$,  then:
$$\lrond({\bf s|D}) \propto \frac{1}{ \sqrt{\det \;
Cov}} \exp [- \frac{1}{2}
                           {\bf D}^{T}Cov^{-1} {\bf D}],$$
where $Cov$ is the covariance matrix
that embodies the cosmological parameter data set ${\bf s}$ and the noise.
If $\lrond$ can be expanded to quadratic order about its maximum, then
the accuracy
with which the  parameters in a given cosmological model can  be
reconstructed
from the data set ${\bf D}$ can be obtained using the Fisher
information matrix $F_{ij}$, whose elements
measure  the width and the shape of the
likelihood function around its maximum
(Bond, Efstathiou \& Tegmark, 1997; Efstathiou \& Bond, 1998):
\begin{eqnarray}
F_{ij}=  \frac{1}{2}tr[A_{i}A_{j}],  \nonumber \;\;\;\;\;
A_{i}=Cov^{-1}({\bf D}) \frac{\partial {\bf D}}{\partial s_{i}}.
\end{eqnarray}
The minimum error  that can be obtained on a parameter
is then given by
\begin{equation}
\delta {\bf s}_{i}=\sqrt{F^{-1}_{ii}},
\end{equation}
depending not only on the experimental
parameters data set,
but also on the target and input cosmological models and the number
of cosmological parameters involved in the computation.
The data set ${\bf D}$ used to construct the Fisher information matrix
can be either a sky map,
the power spectrum, the cross-correlation function, or  some topological
descriptor like genus.

\subsection{Power spectrum statistics}
The most common way to compute cosmological
parameters is to use the power spectrum estimates
(see e.g. Bartlett et al.,1998b; Efstathiou \& Bond, 1998; Zaldarriaga
1998).
This method still has several difficulties
that could lead to biased estimates of the  parameters (Wandelt, Hivon \&
Gorski, 1998b; Bartlett et al., 1998a,b).
They arise
mainly because Galactic cuts in the map,  non-uniform sky coverage,
anisotropic noise and
systematic effects (like those induced by the foregrounds
subtraction) make $C_{l}$ non-Gaussian correlated quantities.
In fact, even in Gaussian theories the $C_{l}$,
that represents the variances of
individual spherical harmonic coefficients, are $\chi^{2}$ distributed.
Hence, the standard likelihood method is not strictly applicable.

If the temperature anisotropy power spectrum alone is used, then the
Fisher information matrix given by equation (23) can be written as
\begin{equation}
F_{ij}=\sum_{l}
       \frac{ \partial C_{Tl} }{ \partial s_{i} } \cdot
        Cov^{-1}({\hat C}^{2}_{Tl}) \cdot
       \frac{ \partial C_{Tl} }{ \partial s_{j} }.
\end{equation}
If  both anisotropy and polarization power spectra are used,
the Fisher information matrix reads as (Zaldarriaga \& Seljak 1997;
Zaldarriaga 1997)
\begin{equation}
F_{ij}=\sum_{l} \sum_{X,Y} \frac{\partial C_{Xl}}{\partial s_{i}}
              Cov^{-1}({\hat C}_{Xl},{\hat C}_{Yl})
                            \frac{\partial C_{Yl}}{\partial s_{j}},
\end{equation}
where $X$ and $Y$ stands for $T$, $E$, $C$ and $B$ power spectra and
$Cov^{-1}$
is the inverse of the covariance matrix. For the purpose of this work
we assume only scalar modes.
Then the relevant covariance matrix elements in Eqs.  (24) and (25) are:
\begin{eqnarray}
& Cov({\hat C}^{2}_{Tl})=\frac{2}{(2l+1)f_{sky}}
       (C_{Tl}+w^{-1}B_{l}^{-2})^{2}, \nonumber%
\\
& Cov({\hat C}^{2}_{El})=\frac{2}{(2l+1)f_{sky}}
                    (C_{El}+w_{P}^{-1}B_{l}^{-2})^{2}, \nonumber%
\\
& Cov({\hat C}^{2}_{Cl})=\frac{2}{(2l+1)f_{sky}}
           [C^{2}_{Cl}+(C_{Tl}+w^{-1}B_{l}^{-2})
                        (C_{El}+w_{P}^{-1}B_{l}^{-2})], \nonumber%
\\
& Cov({\hat C}_{Tl}{\hat C}_{El}
)=\frac{2}{(2l+1)f_{sky}}C^{2}_{Cl},\nonumber%
\\
& Cov({\hat C}_{Tl}{\hat C}_{Cl} )=
     \frac{2}{(2l+1)f_{sky}}
                    C_{Cl}(C_{Tl}+w^{-1}B_{l}^{-2}), \nonumber%
\\
& Cov({\hat C}_{El}{\hat C}_{Cl} )=\frac{2}{(2l+1)f_{sky}}
                    C_{Cl}(C_{El}+w_{P}^{-1}B_{l}^{-2}).%
\end{eqnarray}
Here we set $w=\sum_{c}w_{c}$ with the sum performed over detector
channels,
$w_{c}=(\sigma_{c,pix} \theta_{c,pix})^{-2}$
(Knox 1995) and  $w_{P}=2w$; also,
$B^{2}_{l}=\sum_{c}B_{cl}^{2}w_{c}/w$
accounts for the beam smearing and  $ B_{cl}^{2}=e^{-l(l+1)/l_{s}^{2}}$
is the Gaussian beam profile,
$l_{s}=\sqrt{8 \ln 2}(\theta_{c})^{-1}_{fwhm}$ and
$f_{sky}$ is the fraction of the sky used in the analysis.

Table 1 lists the experimental parameters of the various experiments
that we considered in our calculations. In the Table the parameter set for
MAP
(Bennet et al., 1996) refers
to the current, updated
configuration. Also, we  label by SPOrt-like the parameter
set  used in Section 4 for the computation
of the confidence regions, that is generally consistent with the
parameter data set of SPOrt-ISS, except for a sensitivity level
somewhat better than currently achieved
(Cortiglioni et al., 1997). For each experiment we only consider the
frequency channels where the cosmological signal is not expected to
be masked by the Galactic foreground.

We assume as target model standard reionized CDM model (srCDM)
($\Omega=1$, $h=0.5$, $\Omega_{b}=0.05$, $n_{s}=1$, $\tau_{ri}=1$,
$x_{e}=1$) and
compute the derivatives of the power spectra with respect to
$n_{s}$, $\tau_{ri}$ and $\Omega_{b}$
using the truncated  Taylor series expansion (Efstathiou \& Bond 1998)
\begin{equation}
C_{l}(s_{i})=C_{l}({\boldmath s_{0}})+  \left( \frac{\partial
C_{l}}{\partial s_{i}}
       \right)  \Delta s_{i}.
\end{equation}
Here ${\bf s}_{0}$ is the parameter data set of the target model
and ${\bf s}$  is the parameter data set of the input models that
differs within  $5\%$ of ${\bf s}_{0}$.

Table 2 presents $1$-$ \sigma$ errors on the estimates of the
relevant cosmological parameters obtained from the anisotropy alone
and anisotropy plus  polarization for the experimental parameters
listed in Table 1 and  few values of $f_{sky}$.  For a correct comparison
with previous work
we should consider that the expected errors depend on the number of
model free parameters (4 in our case). Larger errors
obviously arise when as many as $\simeq 10$ free parameters are fitted
(Jungman
et al., 1996a, 1996b).
\subsection{Genus statistics}
The expectation value of  genus
depends on the coherence angle  of the field $\theta_{c}= 2^{1/2} \sigma_0
/ \sigma_1$
(Gott et al., 1990), which can be written as:
\begin{equation}
\theta_{c}^{2}=2\frac{\sum_{l}(2l+1)(C_{l}B_{l}^{2}+w^{-1})}
                     {\sum_{l}l(l+1)(2l+1)(C_{l}B_{l}^{2}+w^{-1})}.
\end{equation}
One should note that the shape of the genus curve  is fixed by the Gaussian
random-%
phase nature of the field and its amplitude depends only the power
spectrum.
For a Gaussian random field  $\theta^{2}_{c}$ is Gaussian
distributed and the standard likelihood method
is fully applicable.

We use the statistics of the coherence angle
to compute the errors on the estimates of the cosmological
parameters for the experimental data sets listed in Table 1.
The Fisher information
matrices can be obtained from Eqs. (24) and (25)
making the following substitutions:
\begin{eqnarray}
& \frac{\partial C_{l}}{\partial s_{i}} \rightarrow
        \frac{\partial \theta^{2}_{c}}{\partial C_{Xl}} \cdot
        \frac{\partial C_{Xl}}{\partial s_{i}},& \nonumber%
\\
& \frac{\partial C_{l}}{\partial s_{j}} \rightarrow
        \frac{\partial \theta^{2}_{c}}{\partial C_{Yl}} \cdot
        \frac{\partial C_{Yl}}{\partial s_{j}},& \nonumber
\end{eqnarray}
\begin{equation}
Cov(C_{Xl}C_{Yl}) \rightarrow
                     \frac{\partial \theta^{2}_{c}}{\partial C_{Xl}}
                     \frac{\partial \theta^{2}_{c}}{\partial C_{Yl}} \cdot
                     Cov(C_{Xl}C_{Yl}) .
\end{equation}
Although the coherence angle does not depends on the
fraction of the sky involved in the analysis,
its  covariance does so through the covariance matrix of the $C_{l}$.

Table 3 lists $1 \sigma$ errors on the estimates of the cosmological
parameters obtained using the coherence angle statistics for few values
of $f_{sky}$.
The inspection of the results listed in Table 2 and Table 3 shows that:
\begin{itemize}
\item Introducing polarization improves the accuracy on
cosmological parameters from both power spectrum statistics and
genus statistics, and as expected, the accuracy on
$\tau_{ri}$ is  improved most, generally by a factor $\sim 3$.
\item  The genus statistics is more sensitive to $n_{s}$ and $\tau_{ri}$
than the power spectrum technique,
while in most cases $\Omega_{b}$ is better determined in the case of power
spectrum.
\item At the largest beamwidth, which cuts off high order harmonics,
the relative performance of genus statistics improves significantly.
\end{itemize}
The second of the above results may seem somewhat strange at first sight.
However, it can be interpreted by observing the topological analyses
weight higher order harmonics more strongly than other current techniques,
so that it is no surprise that the high-likelihood elongated  ``hills''
in parameter space  usually
have somewhat different slopes (Fabbri \& Torres, 1996). Thus couples of
mutually anticorrelated
parameters may have larger and smaller errors, respectively, for
geometrical reasons. The third result, too, can probably be interpreted in
terms of
the weight given to   high order harmonics, which contrasts the loss of
efficiency
arising from the largest beamwidth.

\section{Conclusion}
The genus technique obviously suffers from parameter degeneracy
problems as any other technique, no matter how sophisticated, that can
be applied to the analysis of CMB data. However, since the anticorrelation
curves are not identical to those arising from other techniques, the joint
utilization of
at least two techniques can
significantly reduce the confidence regions of cosmological parameters.
In this connection,  CMB  polarization is useful in order to provide
independent
data sets.

A topological analysis of polarization maps can contribute to
obtain accurate determination of some cosmological parameters
 that determine the spectral amplitude, in particular $n_s$ and $z_{ri}$.
Moreover it seems especially appropriate for large-beam experiments
which are affected by the suppression of
high order harmonics.

\section*{Acknowledgements}

This work is supported by MCT Grant 3005 AA4031, by the Italian Space
Agency
(ASI) and the Italian Ministry for University and Scientific and
Technological Research (MURST).

\newpage

\begin{center}
Table 1. The experimental parameters.

\small
\begin{tabular}{ccccc}
\hline   \hline
 & $\nu$(GHz) & $\theta_{fwhm}$ &
                $\sigma_{pix}/10^{-6}$ & $w^{-1}_{c}/10^{-15}$ \\ \hline
MAP &60&21'&12.1&5.4 \\
(Bennet et al., 1996)     & 90&12.6'&25.5&6.8 \\ \hline
Planck LFI  & 70& 14'&3.6&0.215\\
(Mandolesi et al., 1998)     & 100&10'&4.3&0.156 \\ \hline
     & 100&10.7'&1.7&0.028\\
PLANCK-HFI&150&8'&2.0&0.022\\
(Puget et al., 1998)        &220&5.5'&4.3&0.047\\ \hline
SPOrt-like & 60&$7^\circ$& 1.0 & 14.15 \\
(Cortiglioni et al., 1997)&90&$7^\circ$ &1.0&14.15     \\ \hline \hline
 \end{tabular}
 \normalsize
 \end{center}

\newpage
Table 2. $1-\sigma$ errors on the estimates of the cosmological parameters
obtained from the power spectrum statistics.

\small
\begin{tabular}{ccccc|ccc}
\hline     \hline
 & &  \multicolumn{3}{c|}{Anisotropy}& \multicolumn{3}{c}{Anisotropy \&
polarization} \\
$f_{sky}$&             & 1.          & 0.7        & 0.5         & 1.
& 0.7         & 0.5 \\ \hline
          & $\delta n_{s} \times10^{3}$&7.21 &8.52 &9.26 &3.19 &3.81 &4.51
\\
MAP      & $\delta \tau_{rec} \times 10^{2}$&4.84 &5.68 &6.92 &2.30 &2.67
&3.09 \\
          & $\delta \Omega_{b}\times
10^{3}$&1.15&1.41&1.62&0.96&1.16&1.30\\ \hline
         & $\delta n_{s} \times 10^{3}$&2.97&3.55&4.21&1.02&1.22&1.44\\
PLANCK-LFI& $\delta \tau_{rec} \times
10^{2}$&2.41&2.88&3.4&0.78&0.94&1.11\\
          & $\delta \Omega_{b}\times10^{3}$&0.39&0.47&0.55&0.28&0.33&0.39\\
\hline
          & $\delta n_{s} \times 10^{3}$&2.79&3.33&3.93&0.75&0.89&1.05\\
PLANCK-HFI& $\delta \tau_{rec}\times
10^{2}$&2.28&2.73&3.23&0.58&0.69&0.82\\
          & $\delta \Omega_{b}\times
10^{3}$&0.38&0.45&0.53&0.15&0.18&0.21\\ \hline
          & $\delta n_{s}$&0.42&0.51&0.61&0.11&0.14&0.17\\
SPOrt-like     & $\delta \tau_{rec}$&2.12&2.51&2.97&0.66&0.80&0.94\\
          & $\delta \Omega_{b}\times10^{2}$&2.89&3.43&4.10&1.22&1.46&1.73\\
\hline \hline
\end{tabular}
\normalsize

\newpage
Table 3. $1-\sigma$ errors on the estimates of the cosmological parameters
from genus statistics.

\small
\begin{tabular}{ccccc|ccc}
\hline     \hline
 & &  \multicolumn{3}{c|}{Anisotropy}& \multicolumn{3}{c}{Anisotropy \&
polarization} \\
$f_{sky}$&             & 1.          & 0.7        & 0.5         & 1.
& 0.7         & 0.5 \\ \hline
       & $\delta n_{s} \times 10^{3}$& 3.74 & 4.56 & 5.27&1.31&1.59&1.82 \\
MAP   & $\delta \tau_{rec} \times 10^{2}$&2.11&2.47&2.95&0.76&0.89&1.02 \\
       & $\delta \Omega_{b}\times 10^{3}$&2.99&3.61&4.27&1.41&1.69&1.99\\
\hline
       & $\delta n_{s}\times 10^{4}$&8.12&9.71&11.49&2.62&3.13&3.70 \\
PLANCK-LFI&$\delta \tau_{rec}\times 10^{3}$&4.93&5.89&6.97&1.59&1.90&2.25\\
       & $\delta \Omega_{b}\times 10^{3}$&0.80&1.08&1.28&0.32&0.38&0.46\\
\hline
       & $\delta n_{s} \times 10^{4}$&6.88&8.22&9.73&2.21&2.64&3.13 \\
PLANCK-HFI&$\delta \tau_{rec} \times 10^{3}$&4.26&5.09&6.03&1.37&1.74&1.95
\\
       &$\delta \Omega_{b}\times 10^{3}$&0.81&0.96&1.14&0.29&0.34&0.41\\
\hline
       &$\delta n_{s}$&0.1&0.12&0.15&0.035&0.042&0.051 \\
SPOrt-like&$\delta \tau_{rec}$&0.3&0.35&0.42&0.096&0.11&0.13\\
           & $\delta \Omega_{b}\times
10^{2}$&1.81&2.18&2.58&0.66&0.78&0.93\\\hline \hline
\end{tabular}
\normalsize

\end{document}